\documentclass[aps,twocolumn, secnumarabic,nobalancelastpage,amsmath,amssymb,nofootinbib]{revtex4-2}

\usepackage[utf8]{inputenc}
\usepackage{graphicx}
\usepackage{bm}% bold math
\usepackage{bbm}
\usepackage{bbold}
\usepackage{amsmath}
\usepackage{braket}
\usepackage[normalem]{ulem}
\usepackage{soul}
\usepackage{txfonts}
\usepackage{tabularx}
\usepackage{tabularray}
\usepackage[usenames,dvipsnames]{color}
\setstcolor{red}

\usepackage{geometry}
\newlength\myboxwidth
\begin{document}

\title{Die Separation for Mitigation of Phonon Bursts in Superconducting Circuits}

\author{Guy Moshel}
\email{gmoshel@campus.technion.ac.il}

\author{Omer Rabinowitz}
\author{Eliya Blumenthal}
\author{Shay Hacohen-Gourgy} 
\affiliation{Department of Physics, Technion - Israel Institute of Technology, Haifa 32000, Israel}
\date{\today }

\begin{abstract}
Cosmic rays and background radioactive decay can deposit significant energy into superconducting quantum circuits on planar chips. This energy converts into pair-breaking phonons that travel across the substrate and generate quasiparticles, leading to correlated energy and phase errors in nearby qubits. To mitigate this, we fabricated two separate dies and placed them adjacently without a galvanic connection between them. This blocks phonon propagation from one die to the other. Using microwave kinetic inductance detectors on both dies, we successfully detected high-energy bursts and conclusively demonstrated the blocking effect. However, we also observed simultaneous events in both dies, likely from a single cosmic particle traversing both dies.
\end{abstract}

\maketitle
It is well-established that impacts of cosmic rays and high-energy particles from background radioactivity on planar superconducting quantum circuits produce high-energy phonon bursts in the substrate~\cite{mcewen2022resolving,harrington2024synchronous,grunhaupt2018loss}. Cosmic particles that reach sea level are mostly muons with energy in the GeV range. These particles easily penetrate kilometers of ground, gradually dispensing their energy before stopping. When they pass through a chip they initiate a burst, which then propagates through the substrate and causes detrimental quasiparticle (QP) poisoning in qubits positioned in the affected areas~\cite{martinis2021saving}. The impacts of background radioactive radiation have a similar effect. Substantial experimental effort was made to characterize the dynamics and consequences of these bursts~\cite{Swenson2010highSpeed,moshel2024propagation,gordon2022environmental,yelton2024modeling,thorbeck2023two}. A major concern is that the events induce correlated errors, which cannot be corrected using standard quantum error correction protocols. Many approaches have been suggested to mitigate these errors, including shielding the device underground~\cite{bertoldo2023cosmic,cardani2021reducing}, adding QP traps to the circuit~\cite{hosseinkhani2017optimal,nsanzineza2014trapping}, adding phonon traps to the substrate~\cite{henriques2019phonon,iaia2022phonon,bargerbos2023mitigation}, gap engineering of the Josephson junction to make the transmon less sensitive to QPs traversal~\cite{mcewen2024resisting}, and post-selection of the non-affected calculations in the final quantum processor~\cite{orrell2021sensor}. However, most of these methods provide only partial suppression of the effect and cannot be implemented in some scenarios. In this work, we investigate a very robust solution, namely the separation of a single large chip into distinct dies with no galvanic connection between them. We show that this approach is highly effective in containing the affected regions in a single die, potentially eliminating the correlated nature of the induced errors. The dies can be placed in high proximity to each other, allowing for strong capacitive coupling between the circuits by using appropriate interconnects~\cite{gold2021entanglement,conner2021superconducting}. This enables the exchange of quantum information between the circuits, without also transferring pair-breaking phonons. \\

\begin{figure}[htp!]
    \centering
    \includegraphics[width=\linewidth, trim=0 0 0 0, clip]{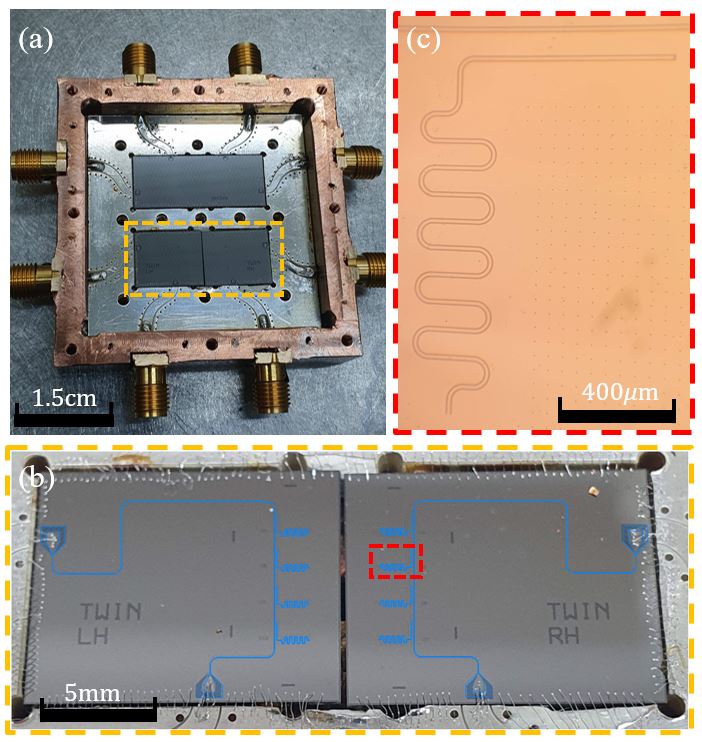}
    \caption{(a) Image of the PCB and copper box on which the experiment dies were mounted, with the lid off. The current experiment was done on the chip in the bottom slot, marked by a yellow dashed line. (b) Close-up image of the mounted dies with wirebonds, superimposed with a schematic drawing of the layout in blue. Each die was held by four pedestals at its corners, with no direct contact between the dies. (c) Micrograph of one of the MKIDs, marked in (b) with a red dashed line.}
    \label{fig:chipLayout}
\end{figure}

Burst energy can potentially cross between dies in the following ways: through the base on which both dies are mounted; through the circuit ground to which both dies are electrically connected; and through pair-breaking radiation emitted from one die and absorbed in the other. In this work, we measured whether a single burst was visible within the same die and between different dies. This can indicate whether physical separation is a viable way to block the propagation of bursts across dies. Bursts were detected using microwave kinetic inductance detectors (MKIDs)~\cite{zmuidzinas2012superconducting} implemented as $\lambda/4$ transmission line resonators. In MKIDs, a substantial fraction of the fundamental electromagnetic mode energy is stored by the kinetic inductance of the material, which depends on the instantaneous density of Cooper pairs (CPs) in the superconductor. It is common to write the resonant frequency of the device as
\begin{equation}
    \omega_0=\frac{1}{\sqrt{C(L_{\mathrm{g}}+L_{\mathrm{k}})}},
\end{equation}
where $C$ is the capacitance of the resonator, $L_{\mathrm{g}}$ is the geometric (electromagnetic) inductance and $L_{\mathrm{k}}$ is the kinetic inductance. When the density of CPs is reduced, which can occur due to a high-energy burst, $L_{\mathrm{k}}$ increases proportionally, causing a decrease in $\omega_0$~\cite{mazin2005microwave,gao2008physics}. By continuously measuring the shift in $\omega_0$ we can know in real-time if a burst had occurred, when it had occurred, and what was the recovery time.\\

The chip layout is shown in Fig.~\ref{fig:chipLayout}(b). Each die had a set of four MKIDs in a hanging configuration on a common feedline. For the fabrication, we first performed an electron beam evaporation of a thin aluminum layer directly on a fresh intrinsic silicon wafer. The layer was then patterned using a Tetramethylammonium hydroxide (TMAH) wet etch process. An aluminum thickness of 11nm was measured using an atomic force microscope after etching. The wafer was then diced to the desired shapes of the dies and the dies were mounted inside an oxygen-free high thermal conductivity (OFHC) copper box~\cite{huang2021microwave}, as shown in Fig.\ref{fig:chipLayout}(a). Each die was supported at its corners by four OFHC copper pedestals, with thermal and mechanical anchoring with ge varnish. The dies had a gap of about 0.2mm between them with no direct contact. Electrical connections between the dies and a designated printed circuit board (PCB), which was soldered to the copper box, were made using aluminum wirebonds. The PCB was plated with a silver layer, which is not a superconductor. The box was sealed with an OFHC copper lid and then mounted inside the mixing stage of a commercial Bluefors dilution cryostat. The measurement signal was the transmission signal through the feedline. It was amplified using a high electron mobility transistor (HEMT) amplifier on the 4K stage of the cryostat. \\

\begin{figure}[htp!]
    \centering
    \includegraphics[width=\linewidth, trim=0 0 0 9, clip]{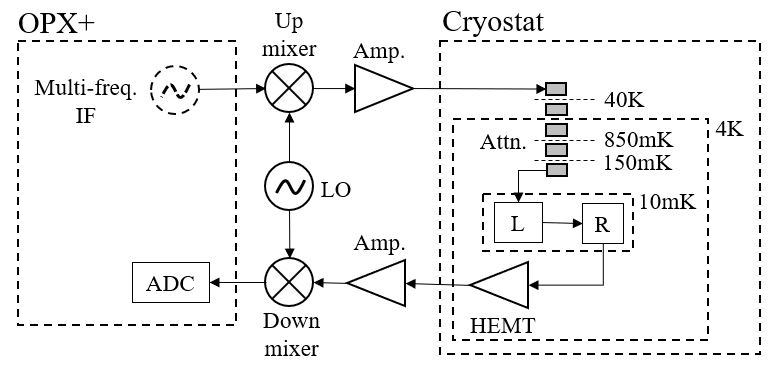}
    \caption{Readout circuit schematic. The two dies were connected in series inside the cryostat.}
    \label{fig:readoutScheme}
\end{figure}

A particle impact anywhere on a die should be detectable in all four MKIDs on the same die due to burst energy propagation via substrate phonons. However, these phonons cannot transfer directly to the other die because of the physical gap. The measurement sequence consisted of simultaneously driving all the MKIDs in their resonance frequencies, thereby loading them with a large number of photons. To increase sensitivity, we used sufficiently high amplitudes that brought the MKIDs outside their linear regimes. after loading, we continuously measured the I and Q quadratures of the MKIDs simultaneously. Every change in the kinetic inductance resulted in a change in the I and Q values, according to the following formula~\cite{probst2015efficient} 
\begin{equation}\label{eq:S21}
    S_{21}(\omega)=1-\frac{\kappa_e}{\kappa_e+\kappa_i+2 i\left(\omega-\omega_0\right)},
\end{equation}
where $\kappa_e$ and $\kappa_i$ are the external and internal linewidths, respectively. When the I and Q values crossed a predefined threshold value, which we determined from the signal fluctuations, a trigger was activated and data was recorded for some specific duration. The MKIDs' drive, continuous readout, and triggered recording was implemented using an OPX+ by Quantum Machines. The measurement setup is presented in Fig.~\ref{fig:readoutScheme}.  As a preliminary step, we characterized the MKIDs in the frequency domain using a network analyzer, PNA-L model N5232B by Keysight. Low power resonance frequencies and linewidths are listed in Table~\ref{table:MKIDsParams}.\\

\begin{table}
\begin{center} 
\caption{Low power parameters of the measured MKIDs. $\omega_0/ 2\pi$ is the resonant frequency and $\kappa_{e}$ and $\kappa_{i}$ are the external and internal linewidths, respectively. D7 was not observed, probably because of an inadvertent short-circuit.}
\begin{tblr}{colspec = {@{}X[l]X[l]X[l]X[l]@{}}}
\hline
\hline
MKID   & $\omega_0/2\pi \;\;\;\;\;$ [GHz] & $\kappa_i/2\pi\;\;\;\;\;\;$ [kHz] & $\kappa_e/2\pi\;\;\;\;$ [kHz] \\
\hline
D1     & 3.28387       & 1.5       & 9.5   \\
D4     & 3.32627       & 2.9       & 8.4   \\
D5     & 3.57318       & 2.1       & 13.0  \\
D6     & 3.61828       & 1.9       & 4.5   \\
D7     & -             & -         & -     \\ 
D8     & 3.48319       & 1.2       & 2.6   \\
D9     & 3.50446       & 1.0       & 3.1   \\
D10    & 3.41655       & 0.8       & 0.8   \\
\hline
\hline
\label{table:MKIDsParams}
\end{tblr}
\end{center}
\end{table}

\begin{figure}[htp!]
    \centering
    \includegraphics[width=\linewidth, trim=0 20 0 0, clip]{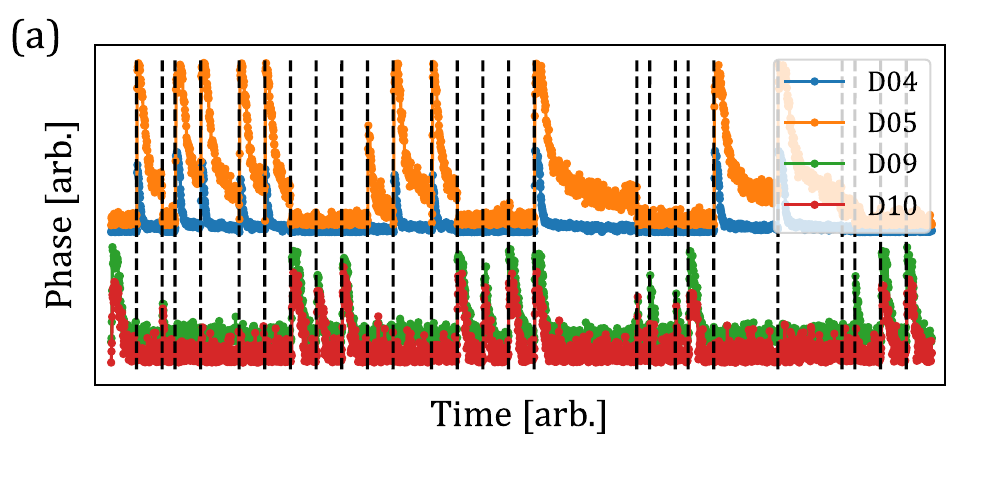}
    \includegraphics[width=\linewidth, trim=0 20 0 0, clip]{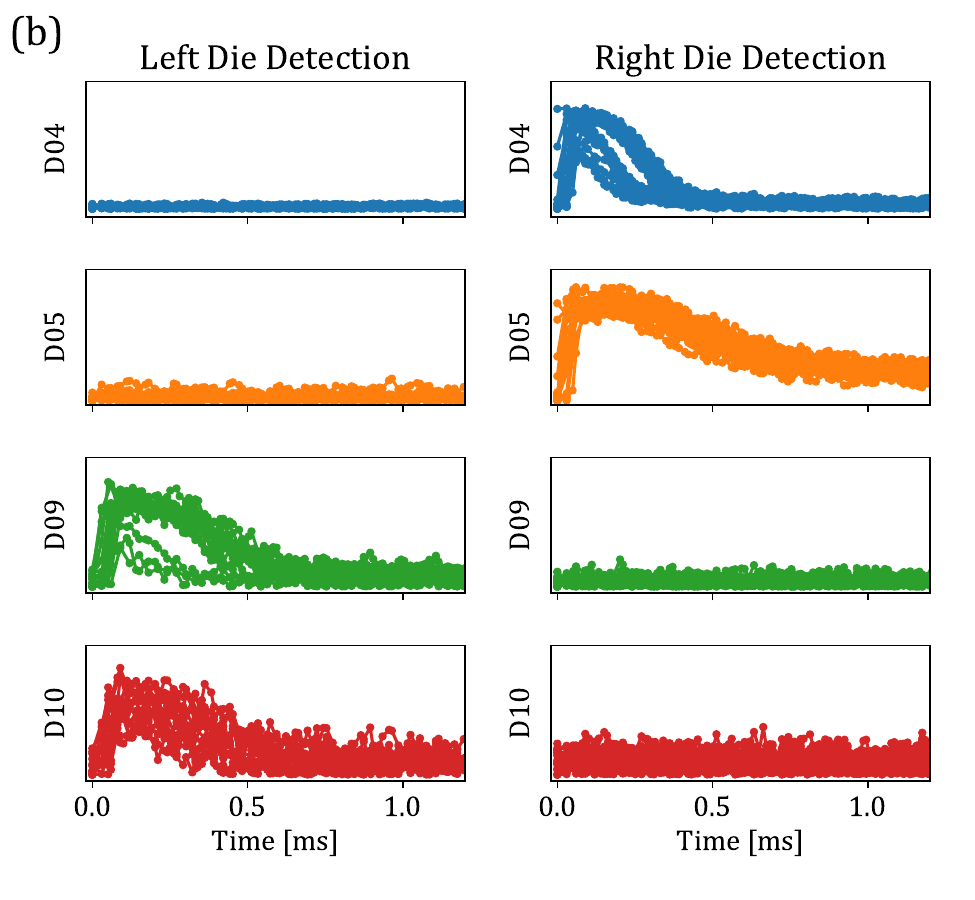}
    \caption{Normalized phases of several typical bursts measured simultaneously in four MKIDs, two from the right die (D04 and D05) and two from the left die (D09 and D10). (a) Bursts displayed in the order of detection. Traces from the right die are artificially shifted above traces from the left die for better visibility. Different events are separated by dashed black lines, which represent an arbitrary time-jump between events. (b) Superimposed bursts showing events detected on the left die (left column) and events detected on the right die (right column).}
    \label{fig:bursts}
\end{figure}

Due to technical limitations, we only measured four MKIDs simultaneously. We chose to measure D04 and D05 on the right die, and D09 and D10 on the left die. Typical time traces of the $S_{21}$ phases of several bursts are presented in Fig.~\ref{fig:bursts}. We present only the phase data because it is the most sensitive and no additional information is gained from the amplitude data. It is evident that the detections are usually in pairs: either in both MKIDs on the right die (D04 and D05) or in both MKIDs on the left die (D09 and D10). This indicates that the propagation of bursts is effectively stopped by the physical separation between the dies. In Fig.~\ref{fig:bursts}(b) several typical superimposed traces show that the recovery time is on the order of 1ms. There is considerable variability in the power of the bursts, which could stem from variations in the impacting particle energy, from variations in the proximity of the impact location to the measured MKIDs, or from variations in the impact angle, which can affect the total energy deposited into the die. 

\begin{figure}[htp!]
    \centering
    \includegraphics[width=\linewidth, trim=0 20 0 0, clip]{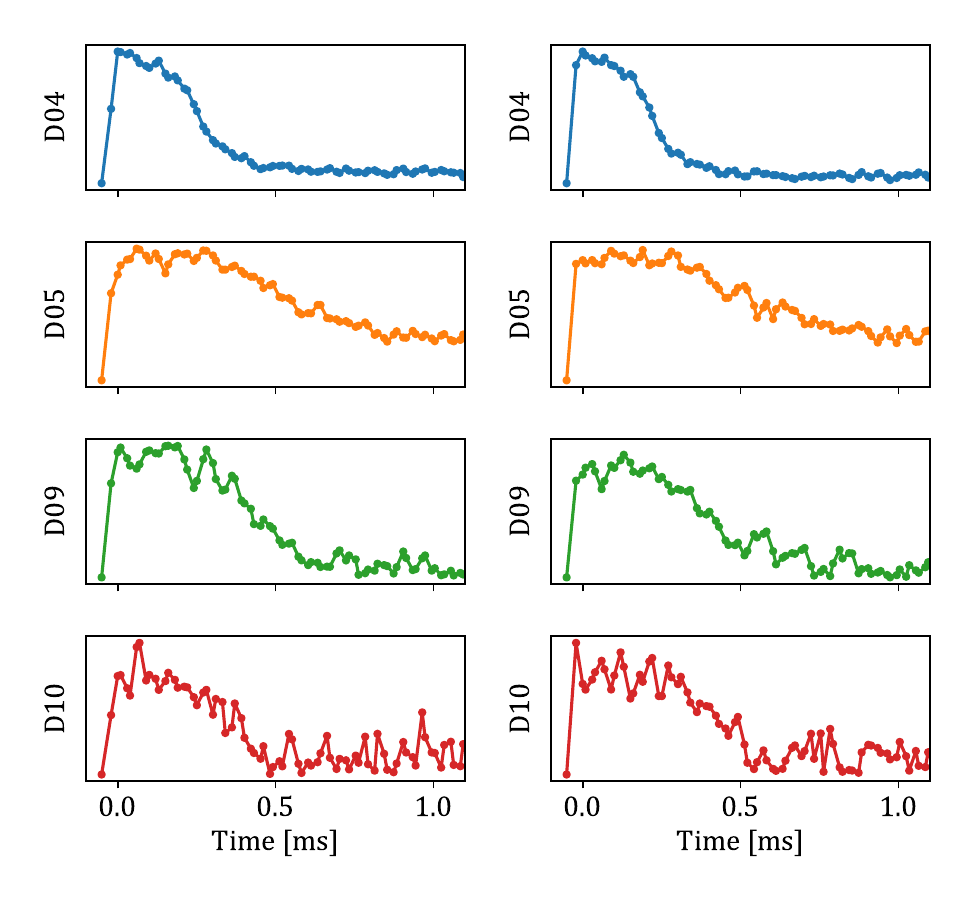}
    \caption{Two examples of four-bursts events, each detected simultaneously in the MKIDs in both dies. The displayed values are normalized phases.}
    \label{fig:burstsAll}
\end{figure}

We also measured events that were detected simultaneously in all four MKIDs, as shown in Fig.~\ref{fig:burstsAll}. We believe these events were caused by
particles that impacted the device at a very shallow angle and traversed both dies directly. This explanation is supported by the typically very high amplitude of these events and by their frequency of occurrence: of the total 352 measured bursts, 10 (2.8\%) were measured simultaneously in all four MKIDs. This frequency matches the expected ratio between single-die events and double-die events due to a single particle impact (4.0\%). Furthermore, we expect the calculated frequency to be an overestimation because the signal generated by some events is small, and they are therefore not measured. The calculation is detailed in the following paragraph. \\

\begin{figure}[htp!]
    \centering
    \includegraphics[width=\linewidth, trim=0 10 0 0, clip]{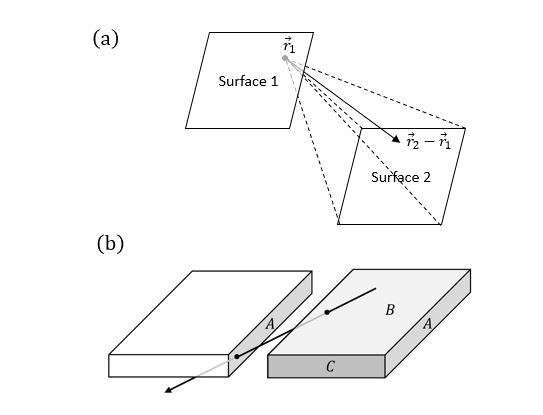}
    \caption{(a) Scheme for calculating the probability of a double impact between two surfaces. (b) A particle trajectory is modeled as a ray that can intersect either or both dies, modeled as rectangular boxes.}
    \label{fig:solid_angle_calc}
\end{figure}

We model the trajectory of a particle as a ray that intersects at least one die. The dies are modeled as perfect rectangular boxes, and we assume that for an event to occur a ray must intersect any of its faces. The model is depicted in Fig.~\ref{fig:solid_angle_calc}(b). Let us denote the surface of entrance of the ray to the first box as surface 1 and the surface of entrance of the ray to the second box as surface 2. The probability to have some double intersection between the two surfaces is
\begin{equation}\label{eq:P}
    P = \int P_1(\vec{r}_1) P_{2|1}(\vec{r}_1) dA_1, 
\end{equation}
where $P_1(\vec{r}_1)$ is the probability of the ray to hit point $\vec{r}_1$ on surface 1, $P_{2|1}(\vec{r}_1)$ is the probability of the ray to hit any point on surface 2 given the previous intersection occurred, and the integration is over the area of surface 1. As a first approximation, we can estimate $P_1(\vec{r}_1)=\text{Const}$, and assume that the total probability to hit surface 1 is proportional to its area, meaning
\begin{equation}
    P_1 = \frac{dA_1}{A_{\text{tot}}},
\end{equation}
with $A_{\text{tot}}$ being the total area of surface 1. $P_{2|1}(\vec{r}_1)$ can be estimated as the solid angle subtended by surface 2 relative to point $\vec{r}_1$, denoted $\Omega_{2|1}(\vec{r}_1)$, divided by $2\pi$. We divide by $2\pi$ and not $4\pi$ because we only count rays coming from one side of surface 1 to avoid double counting when adding contributions from the other surfaces of the box. This is shown schematically in Fig.~\ref{fig:solid_angle_calc}(a).
\begin{equation}
    P_{2|1}(\vec{r}_1) = \frac{\Omega_{2|1}(\vec{r}_1)}{2\pi}=\frac{1}{2\pi}\int \frac{\vec{n}_2 \cdot (\vec{r}_2-\vec{r}_1)}{|\vec{r}_2-\vec{r}_1|^3}dA_2,
\end{equation}
were $\vec{r}_2$ is some point on surface 2 which we integrate over, and $\vec{n}_2$ is a unit vector perpendicular to $dA_2$. After calculating $P$ from Eq.~\ref{eq:P} for a single pair of surfaces, we need to add the contributions of all possible pairs. Because of the geometry of the problem, surface 2 must always be face $A$ in Fig.~\ref{fig:solid_angle_calc}(b), and surface 1 has only five options: face $A$, face $B$ and its opposite, or face $C$ and its opposite. Finally, we multiply the result by 2 to account for rays propagating in the opposite direction.
Notice that our calculation assumes an isotropic distribution of incoming rays, but in practice the distribution of cosmic rays is proportional to $\cos{\theta}$, where $\theta$ is the zenith angle~\cite{grieder2001cosmic}. This should be taken into account in a more accurate calculation. \\

In conclusion, we used MKIDs to demonstrate that high-energy bursts caused by impacts of cosmic rays or background radioactivity affect a large area of a several millimeters of a superconducting circuit fabricated on a silicon chip. We have shown that the propagation of the bursts can be reduced by separating the chip into distinct dies that have a small physical separation between them. We have ruled out the significance of potential energy transfer mechanisms between the dies such as energy transfer through radiation bouncing inside the sealed copper box in which the dies were mounted, and also phonon propagation through the bulk of the copper box and through the PCB to which the dies were electrically connected. Our findings support the use of this method to block the propagation of bursts across a large superconducting circuit, and contain its effect to a predefined area where it can be managed. However, we also observed numerous events where a strong burst was detected simultaneously in both dies. A plausible explanation for these events is that a single high-energy particle passed through both dies. This must be taken into consideration when utilizing the method. In any case, it can be useful to add MKIDs to superconducting circuits as "spectator resonators" that check in real time whether a burst has occurred or not. Then methods such as post-selection or appropriate quantum error correction can be applied to manage the burst. For these applications, a short detection time is paramount. The detection time can be reduced by increasing the signal-to-noise ratio (SNR) of the measurement, enabling a shorter integration time, and also by decreasing the response time of the resonator (increasing the total $\kappa$). This can be done by making the MKID more over-coupled while keeping the total $\kappa$ small enough. From our results, it seems reasonable to obtain a detection time well below $1\mu s$ in an optimized system, which should be sufficient for the aforementioned applications.\\

This research was supported by the Israel Science
Foundation (ISF), the Israel Ministry of Innovation, Science and Technology (MOST), and the Technion’s Helen Diller Quantum Center. The data that support the findings of this study are available from the corresponding author upon reasonable request.

\bibliographystyle{apsrev4-2}
\bibliography{bib}

% \end{center}
%%%%%%%%%% Merge with supplemental materials %%%%%%%%%%
%%%%%%%%%% Prefix a "S" to all equations, figures, tables and reset the counter %%%%%%%%%%
\setcounter{equation}{0}
\setcounter{figure}{0}
\setcounter{table}{0}
\setcounter{page}{1}
\makeatletter
\renewcommand{\theequation}{S\arabic{equation}}
\renewcommand{\thefigure}{S\arabic{figure}}
\renewcommand{\bibnumfmt}[1]{[S#1]}
\renewcommand{\citenumfont}[1]{S#1}
%%%%%%%%%% Prefix a "S" to all equations, figures, tables and reset the counter %%%%%%%%%%

\onecolumngrid
% \subsection*{Device parameters}

\end{document}